\documentclass[aps,amssymb,amsmath, twocolumn, pra, superscriptaddress]{revtex4}
\usepackage{graphicx}
\usepackage{epsfig}
\usepackage{dcolumn}
\usepackage{bm}
\usepackage{bbm}
\usepackage{hyperref}
\usepackage{color}
\usepackage[up]{subfigure}
\newcommand{\be}{\begin{equation}}
\newcommand{\ee}{\end{equation}}
\newcommand{\bc}{\begin{center}}
\newcommand{\ec}{\end{center}}
\newcommand{\bea}{\begin{eqnarray}}
\newcommand{\eea}{\end{eqnarray}}
\newcommand{\ba}{\begin{array}}
\newcommand{\ea}{\end{array}}
\begin{document}

\title{Parrondo's game using a discrete-time quantum walk}
\author{C. M. Chandrashekar}
\email{chandru@imsc.res.in}
\affiliation{Center for Quantum Sciences,
The Institute of Mathematical Sciences, Chennai 600113, India}
\author{Subhashish Banerjee}
\affiliation{Chennai Mathematical Institute, Padur PO, Siruseri 603103, India}
\affiliation{Indian Institute of Technology Rajasthan, Jodhpur 342011, India}


\begin{abstract}

We present a new form of a Parrondo game using discrete-time quantum walk  on a line.
The two players $A$ and $B$ with different quantum coins operators, individually losing the game 
can develop a strategy to emerge as joint winners by using their coins alternatively, or in combination for each step of the quantum walk evolution. We also present a strategy for a player $A$ ($B$) to have a winning probability more than  player $B$ ($A$). Significance of the game strategy in information theory and physical applications are also discussed.
\end{abstract}

\maketitle
\preprint{Version}



\section{Introduction}
\label{intro}

Game theory gives an account of how the involved parties  decide  strategies in their personal interest in a given situation,  on rational grounds. It has found to have relevance to social sciences, biology, and economics \cite{NM47, Dav70, Axe84, Mye91, Pou92}, among others. By replacing classical probabilities with  quantum amplitudes and allowing the players to employ superposition, entanglement and interference, quantum game theory has lead to  new interesting effects and has become an active area of research. Quantum game can be quantified as, any  quantum system which can be manipulated by two or more parties according to their personal interest  \cite{EWL99, EW00}.  
\par
The system in which we propose a quantum game in this Letter is a quantized version of the classical random walk (CRW), that is, the quantum walk system. Quantum walk (QW) evolution on a particle involves the quantum features of interference and superposition, resulting in the quadratically faster spread in position space in comparison to its classical counterpart, CRW  \cite{Ria58, FH65, ADZ93,  DM96, FG98}. QWs  are  studied  in   two  forms: continuous-time  QW  (CTQW) \cite{FG98}  and  discrete-time QW  (DTQW) \cite{ADZ93, DM96, ABN01, NV01}
and are found to be very useful from the perspective of quantum algorithms \cite{Amb03, CCD03, SKB03, AKR05, HRB10}; to demonstrate  coherent  quantum control over atoms; quantum phase transition \cite{CL08}; to explain phenomena such
as the  breakdown of an electric-field driven  system \cite{OKA05} and direct  experimental  evidence  for  wavelike energy  transfer  within photosynthetic systems \cite{ECR07, MRL08};  to generate entanglement between spatially separated systems \cite{CGS10}; to induce dynamic localization in Bose-Einstein condensate in an optical lattice \cite{Cha11a}. Experimental implementation of  QWs has been made in  an NMR system \cite{DLX03, RLB05, LZZ10}; in the continuous tunneling of light fields  through waveguide lattices \cite{PLP08}; in  the  phase space  of trapped  ions  \cite{SMS09, ZKG10}; with single optically trapped  neutral atoms  \cite{KFC09}; and  with  single photons  \cite{SCP10, BFL10}. Various  other  schemes have  been proposed for  their  realization in different  physical systems \cite{RKB02, EMB05, Cha06}.
\par
In this Letter we present a quantum game, in the form of Parrondo's game, 
using a single-particle QW system. Parrondo's game involves two games, which when played individually, produce a losing expectation and when played in any alternating order, winning expectation is produced \cite{MB02a, HAT00, AAT04}.  Parrondo's games were devised, originally, to provide a mechanism to harness Brownian motion and convert it to directed motion, or more generally, a Brownian motor, without the use of macroscopic forces or gradients \cite{Rei02}.     They have since found applications in many areas. Their applications have also been made in the quantum regime. Grover's algorithm \cite{LG96} has been analyzed in the context of quantum Parrondo's paradox \cite{LJ02}.  A  quantum implementation of a capital-dependent Parrondo's paradox, using an economical number of qubits, has been presented in \cite{GM05}. Different forms of Parrondo's game using a QW with a space and time-dependent potential \cite{MB02b},  (with noise \cite{Mey03}), using a QW with history dependence \cite{FAJ04}, using a QW as a source of randomness for randomly switched strategies \cite{KMB07} and using cooperative QW in
which interaction between multiple participants replaces position-dependence as an opportunity
for the Parrondo effect to occur  \cite{BFT08}, have been introduced earlier. 
We present a simple form of Parrondo's game using two players $A$ and $B$ with different quantum coins as quantum coin operators to evolve single-particle DTQW on a line. If the player $A$ ($B$), using his coin evolves the DTQW to $t$ steps such that the probability on the right (left) hand side of the origin ($x=0$) of the particle subjected to QW is greater than that on the left (right), the player $A$ ($B$) is declared as winner. In the game form we present, the players $A$ and $B$ individually losing the game using their coins can develop a strategy 
to maintain equilibrium (equal probability on both the sides of the origin) and emerge as joint winners using their coins alternatively, or in combination for each step of the DTQW evolution. We also present a situation where player $A$ ($B$) can have a winning probability more than a player $B$ ($A$) and emerge as a solo winner. Recently, using the chirality distribution of the DTQW, a coin flipping game has been presented \cite{RH10}. We believe that the manipulation of DTQW in the form of game using multiple players will give a general framework for application of QW, using multiple coins, to algorithms and various physical processes.
\par
In Sections \ref{dtqw} and \ref{pg}, we present the standard form of DTQW and Parrondo's game. In Section \ref{pgqw}, a game using DTQW, in the form of Parrondo's game, is presented. Different strategies for different situations is discussed in Section \ref{ws}. In Section \ref{conc},  we conclude by discussing  the significance of game strategy in information theory and physical applications. 

\section{Discrete-time quantum walk }
\label{dtqw}

The DTQW  in  one-dimension is  modelled  as  a  $2\times K$ system, that is,   a two-level system  (a qubit) in the  Hilbert space  ${\cal H}_c$, spanned  by $|0\rangle$  and  $|1\rangle$, and  a  $K$ level position system, a position degree  of freedom in  the   Hilbert  space  ${\cal  H}_p$,  spanned by $|\psi_x\rangle$, where $x \in {\mathbbm  I}$, the set of integers. 
A  $t$-step DTQW, with unit time required for each step of walk, is generated by  iteratively applying  a unitary  operation $W$  which acts  on the Hilbert space 
${\cal H}_c\otimes    {\cal     H}_p$:    
\be
|\Psi_t\rangle=W^t|\Psi_{\rm ins}\rangle.
\ee  
Here  $|\Psi_{\rm ins}\rangle  = (\cos(\delta        /2)|0\rangle+       \sin(\delta/2)       e^{i\phi}
|1\rangle)\otimes |\psi_0\rangle$,
is an arbitrary initial state of the particle at the initial position $x=0$ and $W\equiv S(B \otimes  {\mathbbm 1})$, where 
\be
\label{coin1}
B = B_{\xi,\theta,\zeta}       \equiv      \left(      \begin{array}{clcr}
  \mbox{~}e^{i\xi}\cos(\theta)      &     &     e^{i\zeta}\sin(\theta)
  \\ e^{-i\zeta} \sin(\theta) & & - e^{-i\xi}\cos(\theta) 
\end{array} \right)\in U(2),
\ee
is  the quantum coin operation which will evolve the particle into the superposition of the particle state. $S$ is a  controlled-shift operation          
 \be            S\equiv         \sum_x \left [  |0\rangle\langle
0|\otimes|\psi_x-1\rangle\langle   \psi_x|   +  |1\rangle\langle
1|\otimes |\psi_x+1\rangle\langle \psi_x| \right ], 
 \ee
and shifts the particle in superposition to superposition of position space.
The probability to find the particle at site $x$ after $t$ steps is given by $P(x,t)  = \langle
\psi_x|{\rm tr}_c (|\Psi_t\rangle\langle\Psi_t|)|\psi_x\rangle$.
\par
For a particle, with initial state at the origin
\be
|\Psi_{\rm ins} \rangle  = \frac{1}{\sqrt 2}(|0\rangle + i |1\rangle) \otimes |\psi_{0}\rangle,
\ee
using an unbiased coin operation, that is, $B_{\xi,\theta,\zeta}$ with $\xi = \zeta = 0$ ($B_{0, \theta, 0} \equiv B_{\theta}$), the variance after $t$ steps of walk is $[1 - \sin(\theta)] t^2$ and a symmetric probability distribution in position space is obtained. Choosing $\theta = 45^{\circ}$ ($B_{\pi/4}$) leads to a standard form of DTQW,  the Hadamard walk \cite{CSL08}. 
For  $\theta=0$, the two states, $|0\rangle$ and $|1\rangle$ move away from each other ballistically without any interference effect. With increase in $\theta$, interference effect is seen and the distribution which is wider for low values of $\theta$, becomes narrower with increase in $\theta$. The interference effect again disappears for the other extreme value of $\theta = \pi/2$.  The two horned peaks on either side of the distribution, which move away with increase in the number of steps, makes QW highly transient in nature.  Small amount of decoherence in QW dynamics
enlarges the range of quantum dynamics by providing a wider range of possibilities for tuning the properties of QWs \cite{Ken07} and to observe symmetries \cite{CSB07}. Aperiodic QW \cite{RMM04} and disordered QW \cite{Cha11a} also enlarge the range of QW dynamics. 

\section{Parrondo's game}
\label{pg}

Standard form of Parrondo's game involves games of chance. Two games, $A$ and $B$, when played individually, produce a losing expectation. An apparently paradoxical situation arises when the two games are played in an alternating order, a winning expectation is produced \cite{HAT00, MB02a, MB02b, AAT04}. The apparent paradox that two losing games $A$ and $B$ can produce a winning outcome when played in an alternating sequence was devised by Parrondo as a pedagogical illustration of the Brownian ratchet \cite{HA99}. However, Parrondo's games have important applications in many physical and biological systems, combining processes lead to counterintuitive dynamics. For example, in control theory, the combination of two unstable systems can cause them to become stable \cite{AA01a}. In the theory of granular flow, drift can occur in a counterintuitive direction \cite{RSP87, Kes97, KK02}. Also, the switching between two transient diffusion processes in random media can form a positive recurrent process \cite{PS92}. 
\par
A brief analysis of Parrondo's paradox can be done by considering two games, $A$ and $B$. Game A can be thought of as a simple coin tossing one, such that  a win (say heads of the coin) occurs with a probability $P$ and a loss (say tails of the coin) occurs with a probability $(1- P)$. The scenario where $A$ is  loosing  could be thought of as a game involving weighted coins.
Game $B$ is more involved in that it is a capital dependent game, with the game strategy dependent upon the initial capital available with the player. It is possible to devise situations wherein both
the games are individually loosing, but a randomized combination of the two results in a winning game. This was shown in \cite{HAT00} using Markov chain theory.

\section{Parrondo's game using DTQW}
\label{pgqw}

Here we present a novel scheme of a game in the setting of a DTQW, viz., one in which two players $A$ and $B$ use different quantum coin operators to walk on a single particle. Multiple coin DTQWs have been studied before \cite{BCA03} and the effect of various coin parameters has been studied by \cite{Kon02}. Both these features have been examined together in the form of Parrondo's game for the first time here, to the best of our knowledge. This could have impact on the studies of directed phenomena such as ratchets on DTQWs. 
\par
\begin{figure}[h]
\begin{center}
\epsfig{figure=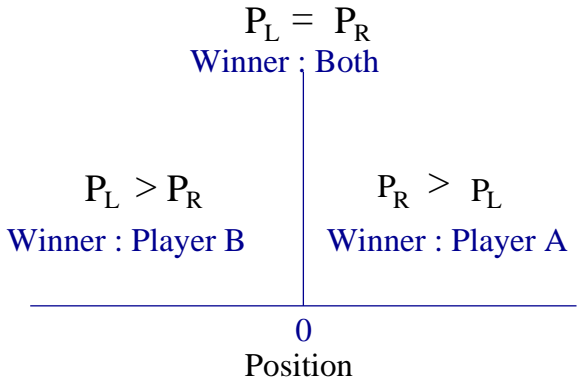, width=8.0cm}
\caption{Pictorial illustration of the conditions for the player to be declared winner.}
\label{fig:qw1}
\end{center}
\end{figure}

To construct a Parando's game like situation using DTQW on a line, we will consider two players $A$ and $B$, as shown pictorially in Fig. (\ref{fig:qw1}),  and 
construct the game as follows, 
{\it 
\begin{itemize}

\item Both the players $A$ and $B$ are given different quantum coin operations $B^{A}_{\xi, \theta, 0}$ and $B^{B}_{0, \theta, \zeta}$ with 
two non-zero variable parameters each and a common shift operator $S$ to evolve DTQW.

\item Initial state of the particle at the origin, $x=0$ on which they have to evolve the walk should be, 
$|\Psi_{\rm ins} \rangle  = \frac{1}{\sqrt 2}(|0\rangle + i |1\rangle) \otimes |\psi_{0}\rangle$.

\item Player $A$ is considered a winner if the probability of finding the particle to the right of the origin, $P_{R}$, is greater than the probability to the left of the origin, $P_{L}$, 
that is, $P_{R}>P_{L}$ after $t$ number of steps of DTQW evolution. Similarly, player $B$ is considered a winner if $P_{L} > P_{R}$ after $t$ number of steps of DTQW evolution. If both the players $A$ and $B$ together manage to maintain equal probability on both the sides, $P_{L}=P_{R}$, they are declared as joint winners.
\end{itemize}
}
\begin{figure}[h]
\begin{center}
\epsfig{figure=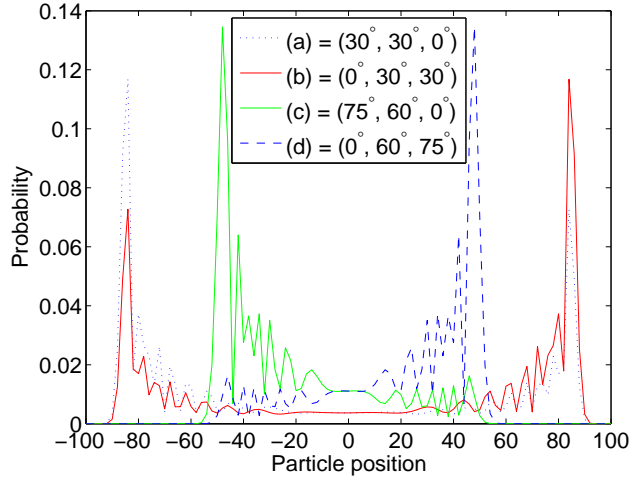, width=9.0cm}
\caption{Spread of probability distribution after implementing 100 steps of DTQW using 
different values for the coin parameters of player $A$,  $B^{A}_{\xi, \theta, 0}$ [(a) and (c)] and player $B$, $B^{B}_{0, \theta, \zeta}$ [(b) and (d)]. Parameter $\xi$ in 
$B^{A}_{\xi, \theta, 0}$ shifts the distribution to the left: (a) = $( \frac{\pi}{6}, \frac{\pi}{6}, 0)$  and (c) = $(\frac{5 \pi}{12}, \frac{\pi}{3}, 0 )$, $P_{L} > P_{R}$. Parameter $\zeta$ in $B^{B}_{\xi, \theta, \zeta}$  shifts the distribution to the right: (b) = $(0, \frac{ \pi}{6}, \frac{\pi}{6})$ and  (d) = $(0, \frac{\pi}{3},  \frac{5 \pi}{12})$, $P_{R} > P_{L}$.  The initial state of the particle used for the walk is $|\Psi_{\rm ins}\rangle = \frac{1}{\sqrt{2}}(|0\rangle + i |1\rangle) \otimes |\psi_{0}\rangle$.}
\label{fig:qw2}
\end{center}
\end{figure}
\noindent From Sec. \ref{dtqw}, we know that $W_{\xi, \theta,  \zeta} = S(B_{\xi, \theta,  \zeta} \otimes {\mathbbm
1})$ on $|\Psi_{\rm ins}\rangle$ implements one step of DTQW. This will evolve the particle to, 
\begin{eqnarray}
\label{eq:condshift2}
W_{\xi, \theta, \zeta}|\Psi_{\rm ins}\rangle =  \frac{1}{\sqrt 2}
 [ \left(e^{i\xi}  \cos(\theta)+ i e^{i\zeta} \sin(\theta)\right )
|0\rangle\otimes |\psi_{-1}\rangle  \nonumber \\
+ \left( +e^{-i\zeta}\sin (\theta) - i e^{-i\xi} 
\cos(\theta)\right) |1\rangle\otimes |\psi_{+1}\rangle ]. 
\end{eqnarray}
The position probability distribution from Eq. (\ref{eq:condshift2}), after the first step, corresponding to
the left and right positions are $\frac{1}{2}[1 \pm \sin(2\theta)\sin(\xi - \zeta)]$.
 These probability distributions would be equal and lead to a left-right symmetry in position, if and only if $\xi = \zeta$. That is, the parameters $\xi \neq \zeta$ introduce asymmetry in the position space probability distribution from the first step itself. The effect of different values of coin parameters after 100 steps of walk is shown in Fig. \ref{fig:qw2} plotted using numerically obtained values. We thus find that the generalized operator $B_{\xi, \theta, \zeta}$ as a quantum coin can bias the probability distribution of the quantum walk in spite of the symmetry of initial state of the particle. This is the key point in developing a winning strategy for  Parrondo's  game using DTQW.
\par
From the preceding analysis, Eq. (\ref{eq:condshift2}) and Fig. \ref{fig:qw2} we know that DTQW with $B_{0, \theta, 0}$ on a particle in state $|\Psi_{\rm ins} \rangle$ results in a symmetric distribution. For $0 < \xi \leq \pi/2$ when $\zeta=0$
using player $A$'s coin $B^{A}_{\xi, \theta, 0}$, the distribution results in asymmetry with $P_{L} > P_{R}$. Therefore, player $A$ can never win a game with the given quantum coin. Similarly, for $0 < \zeta \leq \pi/2$ when $\xi=0$ using player $B$'s coin $B^{B}_{0, \theta, \zeta}$, the distribution results in asymmetry with $P_{R} > P_{L}$.  With increase in $\xi$ and $\zeta$, the difference between $P_{R}$ and $P_{L}$ also increases.
\par
{\it Therefore, both players $A$ and $B$ can never win a game independently with the given quantum coins $B^{A}_{\xi, \theta, 0}$ and $B^{B}_{0, \theta, \zeta}$. }
\par
We should note that even though coin $B^{A}_{\xi, \theta, 0}$ or $B^{B}_{0, \theta, \zeta}$ contribute for asymmetry, the variance of the walk does not deviate much from the variance of the symmetric distribution \cite{CSL08}.  This is because the variance of the walk largely depend on the value of $\theta$ and the value of $\theta$ is same for both the players. This is a genuine game and results from this are not any convention dependent. 

\begin{figure}[h]
\begin{center}
\epsfig{figure=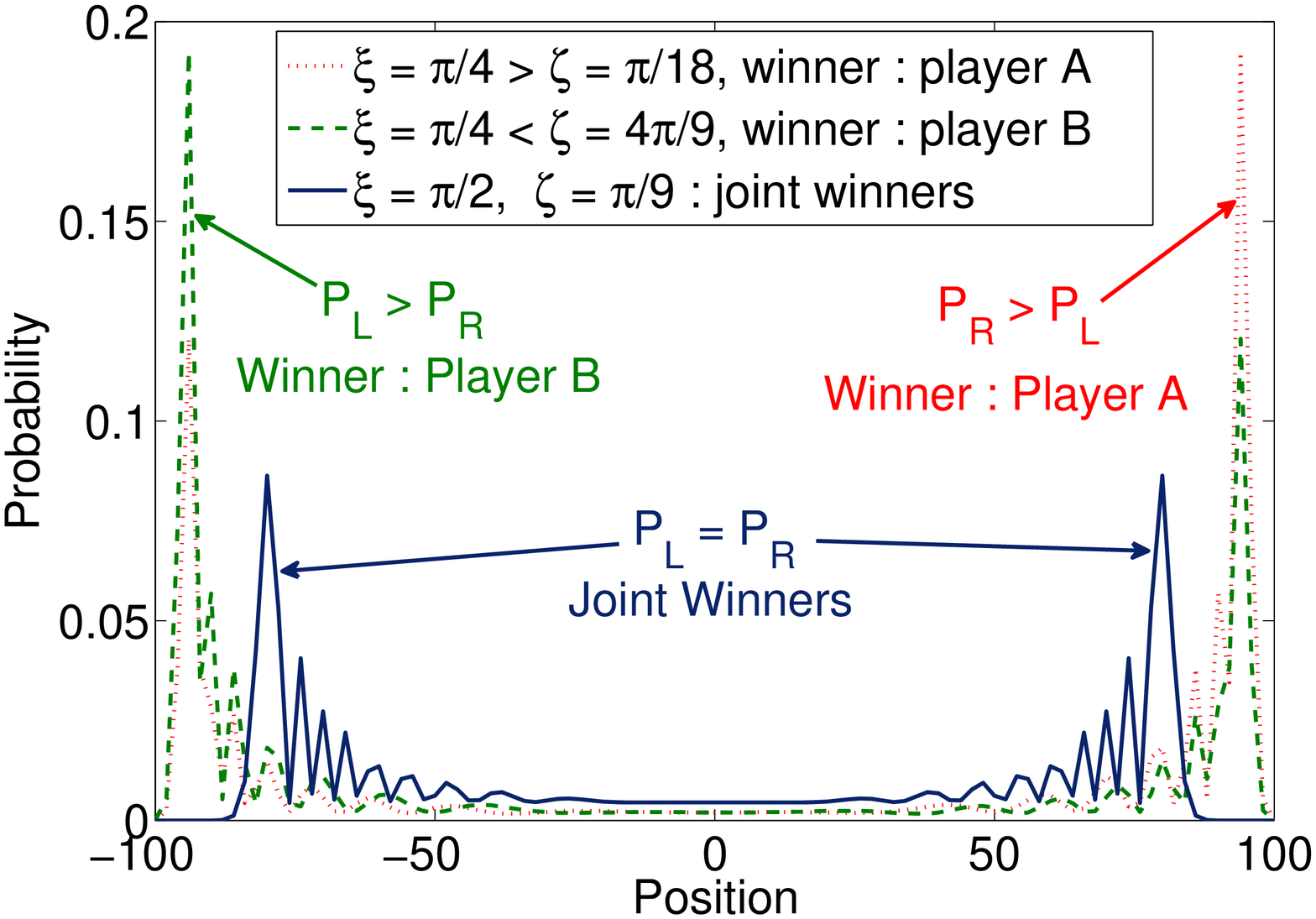, width=9.0cm}
\caption{Spread of probability distribution after implementing 100 steps of DTQW using 
different values for the parameter using operation $W^{BA}_{\xi, \theta, \zeta}$, that is,
player $A$ and $B$ using both of their coins for each step of the walk. If player $A$ chooses $\xi < \pi/2$, player $B$ can emerge as solo winner if he chooses $\zeta > \xi$. If player $B$ chooses $\zeta < \xi$, player $A$ will emerge as solo winner.  However, joint winning can be ensured if player $A$ chooses $\xi = \pi/2$ irrespective of any parameter player $B$ chooses for his coin operation.   The initial state of the particle used for the walk is $|\Psi_{\rm ins}\rangle = \frac{1}{\sqrt{2}}(|0\rangle + i |1\rangle) \otimes |\psi_{0}\rangle$.}
\label{fig:qw3}
\end{center}
\end{figure}

\section{Winning strategy}
\label{ws}
Here we discuss our simple scheme of the winning strategies of the game on a DTQW. The winning strategies are developed making use of the different coin operators. The parameter $\xi$, $\theta$ and $\zeta$ are physically realizable rotations on the two state system \cite{CSL08} and hence restrictions on the rotational degree of freedom lead to situations presented here. We believe this to be a new and simple way of implementing a game on a DTQW.
\par
For the game presented in Sec. \ref{pgqw}, players emerging as joint winners, that is, with $P_{L}=P_{R}$ will be a ideal situation. To emerge as joint winners, the players have to first coordinate between them. They can coordinate at two levels, first level coordination is by alternatively using their coins for each step of the walk, if they are sure that the winner is decided after even number of steps of walk evolution or by using both of their coins, one after the other for every step of the walk. Second level of coordination is to consult each other in choosing the quantum coin parameters $\xi$ and $\zeta$ ($\theta$ is common for both). First level of coordination is a necessary condition for the players to emerge as winners and hence they are always given that choice. Players are not always allowed to coordinate at the second level but they can still emerge as winners. 
\par
First we will discuss the strategy when both the players  agree to cooperate at both the levels, followed by the strategy when they have allowed only for first level of coordination. 

\subsection{Players $A$ and $B$ are allowed to consult each other and the winner is decided after even number of steps of walk evolution}

A simple strategy for players $A$ and $B$ will be to cooperate between themselves for  choosing the coin parameter and emerge as joint winners. This can be achieved if both the players agree,
\begin{enumerate}
\item To use same value for $\xi$ and $\zeta$ in their respective coins; 

\item Use their coins for every alternative step such that their coins are equally used. 
\end{enumerate}
The evolution can be written as,
\begin{eqnarray}
\label{qgame}
W^{B}_{0, \theta, \zeta}W^{A}_{\xi, \theta, 0}.......W^{B}_{0, \theta, \zeta}W^{A}_{\xi, \theta, 0}|\Psi_{\rm ins}\rangle, \\
\mbox{ or~~~~~~~~~~~~~~~~~~ } \nonumber \\
W^{A}_{0, \theta, \zeta}W^{B}_{\xi, \theta, 0}.......W^{A}_{0, \theta, \zeta}W^{B}_{\xi, \theta, 0}|\Psi_{\rm ins}\rangle,
\end{eqnarray}
where
$W^{A}_{\xi, \theta, 0} = S(B^{A}_{\xi , \theta, 0} \otimes {\mathbbm
1})$ and $W^{B}_{0, \theta,  \zeta} = S(B^{B}_{0 , \theta, \zeta} \otimes {\mathbbm
1})$.\\
If the number of steps is odd, the player who uses his coin operation one time more than the other will narrowly lose the game. We will discuss the strategy for a walk with odd number of steps later.

\subsection{Players $A$ and $B$ are not allowed to consult each other and the winner is decided after even number of steps of walk evolution}

When both the players are not allowed to consult each other to choose the coin parameters $\xi$ and $\zeta$, a loosing situation may arise for both the players. To illustrate this further, if player $A$ chooses some value for parameter $\xi$, the winner of the game depends on the choice of parameter $\zeta$ by player $B$. If player $B$  chooses parameter $\zeta < \xi$, player $B$ will emerge as a winner over the player $A$ and if player $B$ chooses $\zeta > \xi$, player $A$ will emerge as a winner over the player $B$. However, a player can make a careful choice of his coin parameter and make sure that he either wins the game alone or emerge as joint winner if the other player is also equally careful. This can be achieved by choosing the lowest non-zero values (initial rule of the game) for the coin parameters $\xi$ and $\zeta$ of the respective players. For example,  player $A$ choosing $\xi = \epsilon$,  $\epsilon$ being the smallest possible value 
greater than zero allowed in the coin operation,  will make sure that player $A$ wins if player $B$ choose $\zeta > \epsilon$ or will emerge as joint winner if player $B$ also choose $\zeta = \epsilon$.

\subsection{When the winner is decided after $t$ (even or odd number) steps of walk}

From the preceding two strategies we know that if the winner is decided after even number of steps of QW, both the players can emerge as joint winners. If the winner is decided after odd number of steps, the player who started the game by using his coin first will narrowly lose the game.
Knowing this, if the players are not told of when the winner will be decided,  none of them will  agree to start the game. In this case they can agree upon a new strategy of using both of their coins for each step of the walk such that both of them would have used their coins equally when the winner is decided.

The evolution can be written as,
\begin{widetext}
\begin{eqnarray}
\label{qgame1}
W^{BA}_{\xi, \theta, \zeta}.......W^{BA}_{\xi, \theta, \zeta}W^{BA}_{\xi, \theta, \zeta}|\Psi_{\rm ins}\rangle, \\
\mbox{ or~~~~~~~~~~~~~~~~~~ } \nonumber \\
W^{AB}_{\xi, \theta, \zeta}.......W^{AB}_{\xi, \theta, \zeta}W^{AB}_{\xi, \theta, \zeta}|\Psi_{\rm ins}\rangle, 
\end{eqnarray}
where
\bea
\label{sta2}
 W^{BA}_{\xi, \theta,  \zeta}  = S(B^{B}_{0 , \theta, \zeta} \otimes {\mathbbm 1})(B^{A}_{\xi , \theta, 0} \otimes {\mathbbm 1})  \equiv   S  \left [  \left(      \begin{array}{clcr}
  \mbox{~}e^{i\xi}\cos^{2}(\theta)+ e^{i\zeta}\sin^{2}(\theta)      &     &     \sin(\theta)\cos(\theta)[1 - e^{i(\zeta - \xi)}] \\
  \sin(\theta)\cos(\theta) [ e^{-i(\zeta-\xi)} -1]    & &  e^{-i\zeta}\sin^{2}(\theta)  + e^{-i\xi}\cos^{2}(\theta) 
\end{array} \right)  \otimes {\mathbbm 1}\right ],
\eea
and 
\bea
\label{sta3}
 W^{AB}_{\xi, \theta,  \zeta} = S(B^{A}_{\xi, \theta, 0} \otimes {\mathbbm 1})(B^{B}_{0 , \theta, \zeta} \otimes {\mathbbm 1})  \equiv   S  \left [  \left(      \begin{array}{clcr}
  \mbox{~}e^{i\xi}\cos^{2}(\theta)+ e^{-i\zeta}\sin^{2}(\theta)      &     &     \sin(\theta)\cos(\theta) [e^{i(\zeta + \xi)} -1]   \\
   \sin(\theta)\cos(\theta) [1 - e^{-i(\xi+\zeta)}]   & &  e^{-i\zeta}\sin^{2}(\theta)  + e^{-i\xi}\cos^{2}(\theta) 
\end{array} \right)  \otimes {\mathbbm 1}\right ].
\eea
\end{widetext}
Action of $W^{BA}$ on $|\Psi_{\rm ins}\rangle$ results in symmetric distribution only for some specific values of $\xi$ and $\zeta$ together.  If player $A$ chooses some $0 < \xi < \pi/2$,  
player $A$ ($B$) can emerge as a solo winner if  player $B$ chooses $\zeta   < \xi$ ($\zeta  > \xi$). 
Following  Eqs. (\ref{sta2}) and (\ref{sta3}), one can guarantee that the player who starts the game can choose a parameter of the coin such that he always wins, either as solo winner or joint winner. For player $A$ starting the game, picking $\xi = \pi/2$ will ensure a joint winning situation even if player $B$ tries to fix $\zeta \neq \xi$ ( $\theta = \pi/4$ is kept the same for both the players).
For player $B$ starting the game, picking $\zeta = \pi/2$ will ensure a joint winning situation even if player $A$ tries to fix $0 <  \xi < \pi/2$ (note that $\xi \neq \pi/2$,  as it will lead to an initial state without shifting to either left or right). In Fig. \ref{fig:qw3}, we have show the winning strategy for walk using $W^{BA}_{\xi, \theta, \zeta}$, Eq. (\ref{sta2}). 

\section{Conclusion}
\label{conc}

Superior performance of quantum strategies is usually seen,  if entanglement is present. In the context of DTQW, this is natural as the walk evolves by entangling the coin and position degrees of freedom. Therefore, making use of the above, and the fact that the walk can be manipulated by varying the parameters in the quantum coin operation, we presented a scheme for a quantum game using DTQW in the form of Parrondo's game. Our system involves two players $A$ and $B$ as quantum coin operators to evolve DTQW on a line.  We presented a situation where both the players are unable to manipulate a walk to an extent to win a game using their coins individually. 
We presented a quantum strategy for the players to cooperate by using their quantum coin operations alternatively and emerge as joint winner for situations where it is conditioned that the winner is decided only after even number of steps of walk evolution.  A different joint winning strategy, by using their coins in combination for each step, is presented for situations where the time (after even or odd number of steps) when the winner will be decided is not known. When both the players are equally careful in choosing their parameters, they can always emerge as joint winners and only a less careful player will be prone to lose the game. Our scheme has fixed a specific initial state for the particle on which the walk is evolved and different coins are allotted to players $A$ and $B$ accordingly. One can choose any initial state for the particle, for example, $|0\rangle$ or $|1\rangle$ or any of its linear combination and different coins and strategies for player $A$ and $B$ can be worked out.  Parameters $\xi, \theta$ and $\zeta$ are physically realizable rotations on the two state quantum system \cite {CSL08, CSB07} and hence restrictions on the rotational degrees of freedom lead to situations presented in this Letter. These strategies, leading the players to emerge as joint or solo winners can be very useful for various physical situations. For example, to arrive at equilibrium (equal on both side of origin), or any non-equilibrium (towards one side) configuration in the probability distribution as required during its application for algorithms or other physical process such as ratchets.    
\par
The causal structure of QW evolution, as shown recently \cite{CBS10},  puts in perspective the nature of quantum information processing involved in a quantum Parrondo's game using QW, in that it provides an upper bound on the propagation of quantum effects of interference and superposition.  The effect of a noisy input to a quantum game has been studied and it is known that a suitable level of noise can enhance the payoff to the players.  In this context,   it  was noted \cite{FA05} in a study of  the effect of decoherence (a form of noise due to interaction of the system with its environment) on a quantum game analogous to a three-player duel, that the boundary in the parameter space, in a transition from the quantum to  classical (by increasing the influence of  noise), changes from convex in the quantum case to linear in the classical one.  This is of interest since convexity is thought to be the basis for Parrondo's paradox.  
\par
As discussed in the introduction, Parrondo's games were devised, originally, to provide a mechanism to harness Brownian motion and convert it to  directed motion. In conclusion,  causal structure of QW evolution, effect of decoherence on Parrondo's game using QW evolution can provide a mechanism to harness QW evolution for various quantum information processing tasks and other applications.

\bibliographystyle{model1a-num-names}

\end{document}